\documentclass{sig-alternate-2013}

\usepackage{times}
\usepackage{helvet}
\usepackage{courier}
\usepackage{graphicx}
\usepackage{color}
\usepackage{soul}
\usepackage{url}
\usepackage{xcolor}
\usepackage{stfloats}
\usepackage{caption}
\usepackage{tabularx}
\usepackage{epstopdf}
\usepackage{rotating}
\usepackage{epstopdf}
\usepackage{siunitx}
\usepackage{booktabs}
\usepackage{subcaption}
\usepackage{tabularx}
\usepackage{subcaption}
\usepackage{array}

\title{Topological Properties and Temporal Dynamics of \\ Place Networks in Urban Environments}

\numberofauthors{4}
\author{
\alignauthor
Anastasios Noulas\\
       \affaddr{Computer Laboratory}\\
       \affaddr{University of Cambridge}\\
\alignauthor
Blake Shaw\\
       \affaddr{Foursquare}\\
       \affaddr{568 Broadway, New York}\\
\and  
\alignauthor
Renaud Lambiotte\\
       \affaddr{Department of Mathematics}\\
       \affaddr{University of Namur}\\
\alignauthor
Cecilia Mascolo\\
       \affaddr{Computer Laboratory}\\
       \affaddr{University of Cambridge}\\
}

\begin{document}
\conferenceinfo{WWW}{'15, May 18--22, 2015, Florence, Italy.} 
\maketitle






\begin{abstract}
Understanding the spatial networks formed by the trajectories of mobile users can be beneficial to applications ranging from epidemiology to local search. Despite the potential for impact in a number of fields, several aspects of human mobility networks
remain largely unexplored due to the lack of large-scale data at a fine spatiotemporal resolution. Using a longitudinal dataset from the location-based service Foursquare, we perform an empirical analysis of the topological properties of place networks and note their resemblance to online social networks in terms of heavy-tailed degree distributions, triadic closure mechanisms and the small world property. Unlike social networks however, place networks present a mixture of connectivity trends in terms of assortativity that are surprisingly similar to those of the web graph. We take advantage of additional semantic information to interpret how nodes that take on functional roles such as `travel hub', or `food spot' behave in these networks. Finally, motivated by the large volume of new links appearing in place networks over time, we formulate the classic link prediction problem in this new domain. We propose a novel variant of gravity models that brings together three essential elements of inter-place connectivity in urban environments: network-level interactions, human mobility dynamics, and geographic distance. We evaluate this model and find it outperforms a number of baseline predictors and supervised learning algorithms on a task of predicting new links in a sample of one hundred popular cities.

\end{abstract}

\section{Introduction}
\label{sec:intro}
Mobile user trajectories are known to exhibit structural
and temporal regularities associated with the daily and weekly cycles of human activity. 
The spatial network formed by user movement, and its topological characteristics
in particular, have been explored in recent research including
the detection of urban neighborhoods~\cite{cranshaw2012livehoods}, place recommendation to mobile users~\cite{noulas2012mining}, touristic route identification~\cite{lucchese2012random}, and a broad range of applications in epidemiology~\cite{bajardi2011human}.

However, the majority of models of human mobility focus exclusively on its spatial characteristics~\cite{brockmann2006scaling, gonzalez2008understanding, simini2012universal}, 
and neglect both network topology and temporal dynamics. 
Recent work has proposed more advanced computational methods that realize some of these aspects, for instance by incorporating information about users' social network~\cite{sadilek2012finding} and their spatiotemporal dynamics~\cite{cho2011friendship}. The applicability of these approaches is limited, however, as they rely on complete knowledge of a user's historic whereabouts 
and social connections as input, which might not be readily available in most domains. 



The goal of this paper is to bridge the gap 
between universal mobility models and complex computational methods in mobility modeling.
As opposed to tracking the whereabouts of individual users, our key idea is to use the aggregate trajectories of users between real-world places to define a network of venues in the city. 
Using a longitudinal dataset from the location-based service Foursquare we empirically analyze \textit{place networks}
in one hundred metropolitan areas across the globe.
Exploiting a set of insights on the growth patterns, temporal dynamics, and topological properties of these place networks,
we then build a new human mobility model that accurately predicts the future interactions between places in urban environments with minimal parameterization and computational costs. Our work is articulated in three parts: 
\vspace{-2.5mm}
\paragraph*{\textbf{Place network growth and temporal pattern analysis}}
We first consider the temporal properties of place networks, and focus on their 
growth over time in terms of edge and node addition processes (Section~\ref{sec:evolution}). In accordance with previous observations in online social networks~\cite{leskovec2005graphs}, we observe a densification pattern, as \textit{the number of edges grows superlinearly to the number of nodes in the system}. 
A saturating effect for node growth is reached quickly nonetheless, when the large majority 
of Foursquare venues have been added to the network. 
\textit{It takes approximately $\mathit{10}$ weeks for mobile users to crowdsource
a large fraction (more than $\mathit{95\%}$) of public places in a city}.
Subsequently, we compare instances of place networks across 
consecutive time windows of observation; we find that \textit{a significant number of new links are generated over time
as users form new spatial trajectories when they navigate between places}. 
The set of places that generate those edges,
however, remains remarkably stable over long periods of time. These results reveal the importance of viewing connections as fleeting entities that emerge dynamically in the network. 
\vspace{-2.5mm}
\paragraph*{\textbf{Topological properties of place networks}}
We then empirically analyze the topological properties of place networks (Section~\ref{sec:analysis}). We make two key observations: 
first, we note that \textit{place networks exhibit the well-known characteristics of social 
networks} such as heavily skewed degree distributions, scale-free properties, small-world behavior, and high clustering coefficients. 
We trace this relationship to the inherent inter-dependence between mobility and social link formation in geographic
space~\cite{sadilek2012finding, cho2011friendship, wang2011human}.
In contrast, we also find a striking difference compared to social networks: 
\textit{they show a resemblance to the web graph presenting a balanced assortative mixing pattern with hub nodes 
connecting to each other but also to low degree nodes.}
This non-social property arises from the different roles played by places in the network. In particular the existence of travel spots (e.g.
train stations or airports), act as intermediate hubs between nearby places (e.g. restaurants, the most frequent place type in 
the network), typically characterized by low degrees. These characteristics are consistent across one hundred cities.
\vspace{-2.5mm}
\paragraph*{\textbf{A new gravity model for link prediction in place networks}}
Finally, the turnover of links in the network over time motivates the following prediction task: given past 
observations about the connectivity of public venues in Foursquare, 
we would like to predict the pairs of places that are likely to connect at a future time (Section~\ref{sec:models}).
Candidate prediction models need to rank highly the pairs of venues 
that are most likely to interact, a task complicated by a number of challenges. 
In particular, the highly volatile, time dependent, link generation process and sparse
data setting may hinder the use of complex prediction algorithms that can be prone to overfitting. The inherently spatial embedding of the network suggests the need for models which integrate appropriately geographic distance as a factor.
 We therefore develop a generalization of \textit{gravity models}~\cite{brockmann2006scaling, gonzalez2008understanding,carrothers1956historical,niedercorn1969economic}, popular in the mobility and transport literature, into which we incorporate the temporal aspects of the system. The model combines information on venue synchronization in terms of user activity, 
in- and out-bound movement towards places, and geographic distance. In practice, it captures the observation that nodes may act as
sources or sinks of users in the course of time, depending on their cycle of activity.
Finally, it incorporates information about the interaction of places on the network level, a valuable aspect 
of attraction that has been ignored by past mobility modeling approaches.
The ranking strategy put forward by the model
outperforms popular supervised learning algorithms by at least two points in the Area Under the Curve (AUC) score, and by a large margin the model adhering to the standard
formulation of gravity in the literature (AUC score 0.905 versus 0.811). This is achieved with minimal requirements for training and optimization, making it ideal in practical application scenarios where expensive
computations can pose a trade-off against the real time demands of many mobile applications. These results are discussed in Section~\ref{sec:evaluation}.

\section{Network Growth and Dynamics}
\label{sec:evolution}
\begin{figure}[t]
        \centering
        \includegraphics[width=0.92\columnwidth]{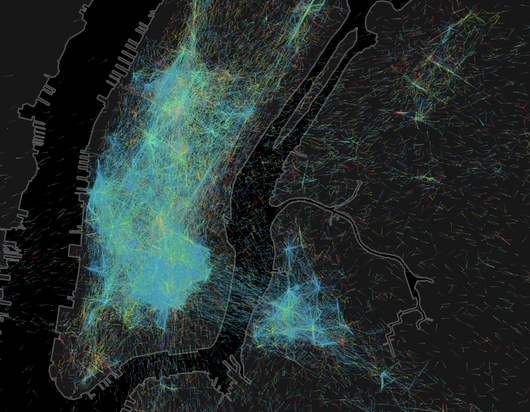}
        \caption{A visualization of the place network for New York City at 11pm.  Each dot represents a user traveling between venues, and is color-coded by the category of the destination with blue being nightlife and green being food. We clearly see the edges of the network formed by people moving between places.}\label{fig:mapnet}
        \vspace{-1.5mm}
\end{figure}

Figure \ref{fig:mapnet} presents a visualization of the place network shaped by the movement of Foursquare users in New York City. 
One can spot hubs being formed at multiple areas of the city, with local transitions connecting them to nearby places 
and occasional long jumps connecting places located further apart from each other, for example when users move between Manhattan and Brooklyn. 

In this section, we investigate the growth patterns of such networks. In particular, we are asking \textit{how does the number of new edges
observed in the network relate to the number of nodes, or places that are being crowdsourced by Foursquare users?} Subsequently, we explore how the network in a city changes over long periods of time, when considering different temporal windows of observation. \textit{Are two temporal network snapshots similar to each other? Or, instead, do mobile users form new trajectories as they explore the city,
contributing towards the formation of a large volume of new links?}

\subsection{Preliminaries on place networks}

We define an urban place network as a \textit{directed} weighted network whose nodes are the popular public places of a city.
For each city, we consider a finite time period of observation $t$ and build a directed graph $G^{t}$ comprised of a set of nodes $V^{t}$ and a set of edges, or links, $E^{t}$. An edge is formed between a pair of places if a Foursquare user directly transitioned between that pair during period $t$. By a direct transition of a user between two places $i$ and $j$, we mean that the user checked in at place $i$ first and his next check-in took place at $j$. If more than one transition occurs between two places, the weight of an edge is incremented accordingly.
We further impose a temporal threshold on each transition so that only direct transitions within $3$ hours are taken into account. This aims to avoid biases related to non-direct user movements. 

\subsection{Network growth patterns}
\label{sec:networkgrowth}
\begin{figure} 
    \centering
    \includegraphics[width=0.78\columnwidth]{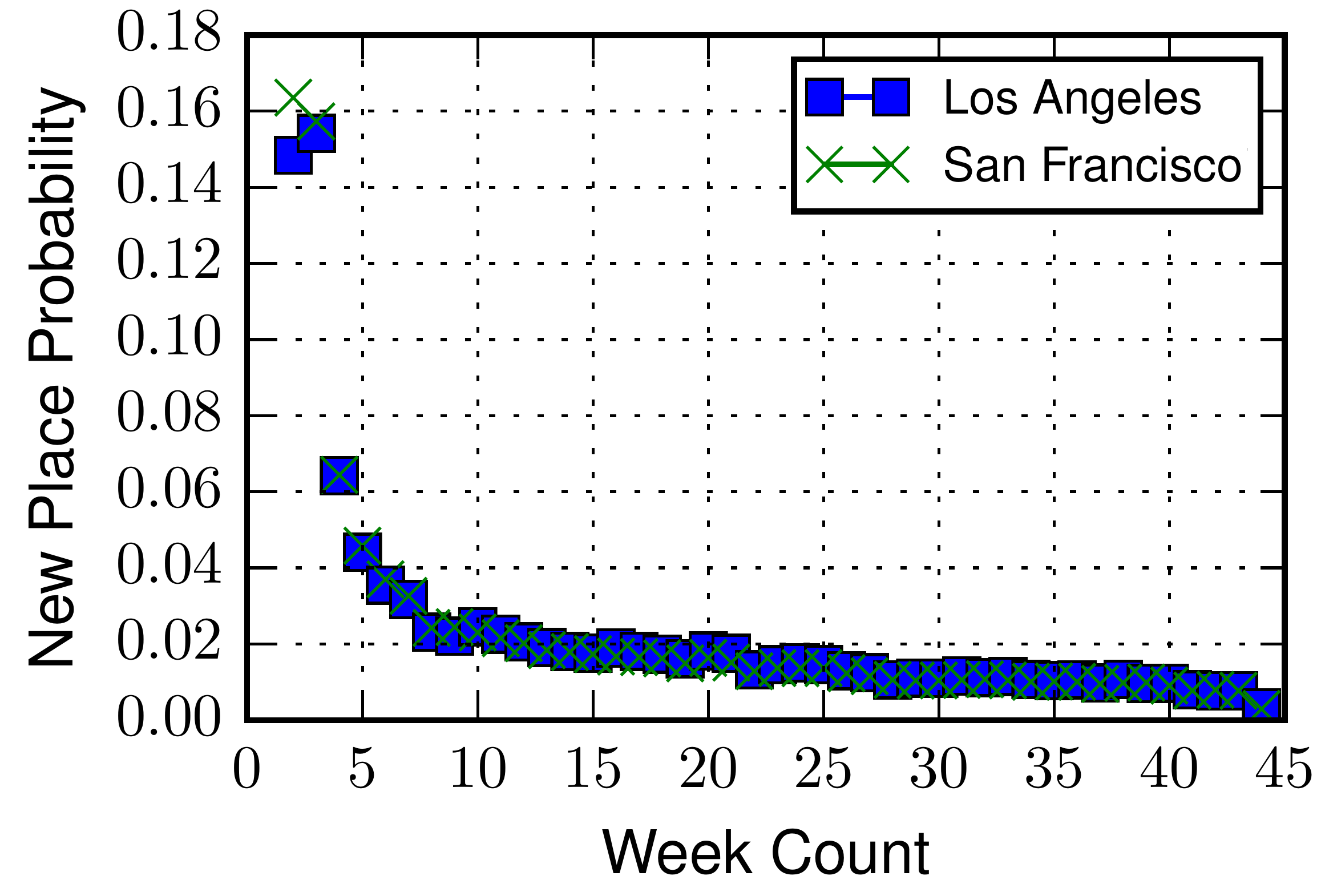}
    \caption{Starting from week $2$, we show
    the fraction of new venues added to the system out of the total observed that week, for Los Angeles and San Francisco.}
    \label{fig:venuegrowth}
\vspace{-1.5mm}
\end{figure}

Network densification is a fundamental phenomenon in network dynamics and relates to the different rhythms with which nodes and edges are added to the network. 
Previous work by Leskovec et al.~\cite{leskovec2005graphs} characterizes empirically the densification process in online social and technological networks showing that the number of edges grows superlinearly with the number of nodes in the network. Specifically, given the number of nodes $n(t)$ observed at a point in time $t$, one is interested in the number of edges $e(t)$ and the way this relationship forms as $t$ grows. Formally we have:
\begin{equation}
e(t) \propto n(t)^\alpha
\end{equation}
Different values of the exponent $\alpha$ imply differences in the expected number of edges over time. A graph with $\alpha = 1$ maintains a stable average degree over time, whereas $\alpha > 1$ corresponds to an increase in the average degree. The findings reported by Leskovec et al. suggest that the latter is the case in many real world networks, and here we investigate whether it holds also in urban place networks. 
\begin{figure}
    \centering
    \includegraphics[width=0.78\columnwidth]{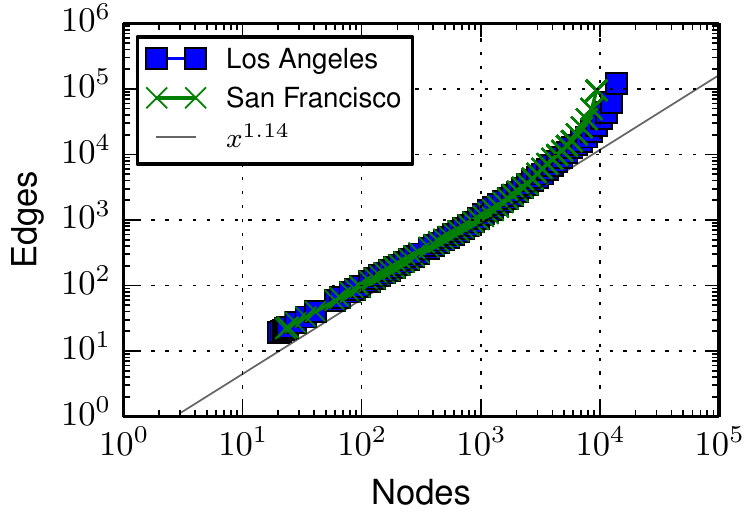}
    \caption{Number of edges versus number of nodes in Los Angeles and San Francisco as the cities become crowdsourced by Foursquare users.}
    \label{fig:growth}
\vspace{-2.5mm}
\end{figure}
We pick a random point in time $t_0$ where we begin monitoring the evolution of a place network and then measure the number of new nodes and edges added by users sequentially.
Figure~\ref{fig:growth} shows the number of edges versus the number of nodes in the cities of Los Angeles and San Francisco, as venue information in these cities becomes crowdsourced over time by Foursquare users. Initially the number of links grows superlinearly with the number of nodes. We have measured using the least squares optimization method an exponent $\alpha = 1.14$ with
a standard deviation $\pm 0.06$ across a set of one hundred cities. However, at a specific city size-dependent threshold, this scaling breaks, as the number of nodes ceases to increase whereas new links continue to appear. At that point, a majority of places have been discovered by the users, and finite-size effects induce a slowing down of new place discovery, as shown in
Figure~\ref{fig:venuegrowth}, where we plot the fraction of newly added venues over time for the two cities. One observes that
it takes approximately $10$ weeks for Foursquare users to crowdsource a significant fraction of the city's set of public venues, as the probability of seeing a new place after the $10$th week drops to approximately $0.02$, with a convergence very close to zero after several weeks. 
One should note that new venues such as retail facilities are continuously created in a city, so the probability may never drop to zero, but the time scale of urban growth is much slower, on the order of months or years~\cite{masucci2013limited}, than the rate of place discovery by Foursquare users.
Unlike an online social network, or other technological networks, that may be able to reach a maximum number of nodes in the order of years (Facebook still adds
new users at a high rate~\cite{fbgrowth}), urban place networks are smaller in size by several orders of magnitude. Even a large metropolis will have on the order of several thousand places, which is a significantly smaller number compared to the hundreds of millions of users
in a social graph.
Importantly, the number of edges crowdsourced by the users remains small as compared to the total number of connections, of the order $n(t)^2$, and no finite-size effect is encountered by link creation, which leads to the patterns of Figure~\ref{fig:growth}.
\subsection{Temporal dynamics of place networks}
\label{sec:networksnaps}
\paragraph*{\textbf{Link generation over time}}
The results of section~\ref{sec:networkgrowth} show that place networks are dynamically evolving entities with new edges being added continuously over time. Given these observations, a natural question to ask is \textit{how do these links persist in time?} Put otherwise, will an observed link
re-appear?

\begin{figure}
        \centering
        \includegraphics[width=0.78\columnwidth]{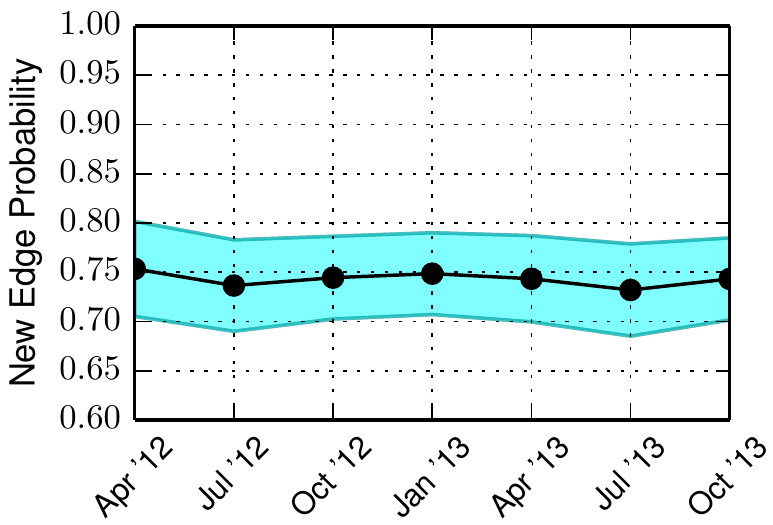}
        \caption{Probability of observing a new edge in the next network snapshot. The probability
        is measured by comparing successive temporal network snapshots. Each point 
        on the x-axis refers to the date that a snapshot ends and its successor begins. The shaded area corresponds to the standard deviations across $100$ cities worldwide.}
        \label{fig:newedgeprob}
\vspace{-1.5mm}
 \end{figure}

Considering three-month temporal snapshots over a period of two years, from the beginning of 2012 to the end of 2014, 
we estimate the probability of a new place, or a new edge, being added in a subsequent time period. Formally, given the set of edges $E^{t}$ observed in a network snapshot during a three month period $t$ and the set of edges $E^{t+1}$ of the subsequent time period, we define the probability of a new edge occurring in the next snapshot as:
\begin{equation}
P_{e} = \frac{|E^{t+1} - E^{t}|}{|E^{t+1}|}
\end{equation}
Figure~\ref{fig:newedgeprob} shows the corresponding probabilities across seven subsequent intersections of network snapshots. The probability of a new edge forming is approximately $70-75\%$ and has small standard deviations across cities as shown by the shaded curve. 

\begin{figure}

        \centering
        \includegraphics[width=0.78\columnwidth]{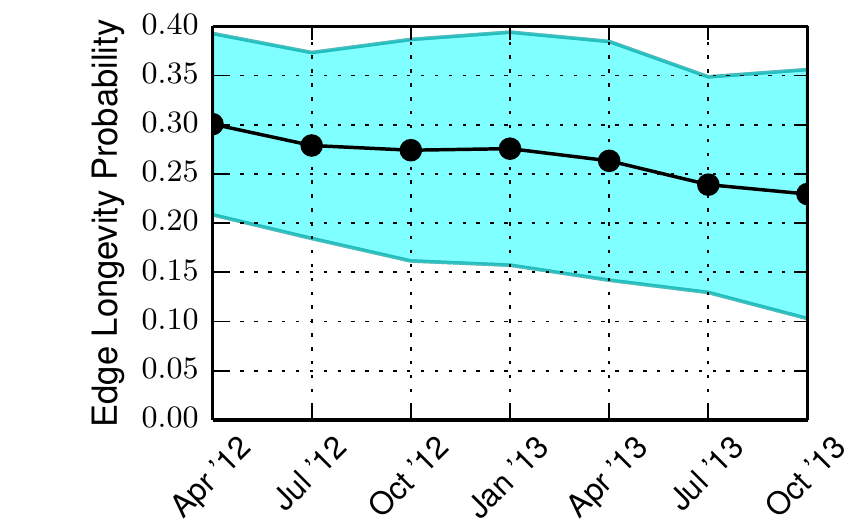}
        \caption{The edge longevity probability measure shows the tendency for an edge to 
        appear in future temporal network snapshots.}
        \label{fig:edgelongevity}
\vspace{-1.5mm}
\end{figure}

A related measurement is the probability that an edge will persist in the network by re-emerging 
consistently in network snapshots over several months. Figure~\ref{fig:edgelongevity} shows the average probability for a given edge's reappearing in snapshots $t+1, t+2 \dots t+n$ given that the edge has appeared in snapshot $t$. We use the term \textit{edge} \textit{longevity} to 
denote this process.
Formally, this probability is defined as the cardinality of the intersection of $n$ consecutive network snapshots, divided by the cardinality of the starting
snapshot $t$:
\begin{equation}
P_{e,n} = \frac{|E^{t} \cap E^{t+1} \dots \cap E^{t+n}|}{|E^{t}|}
\end{equation}

\begin{figure}
        \centering
        \includegraphics[width=0.695\columnwidth]{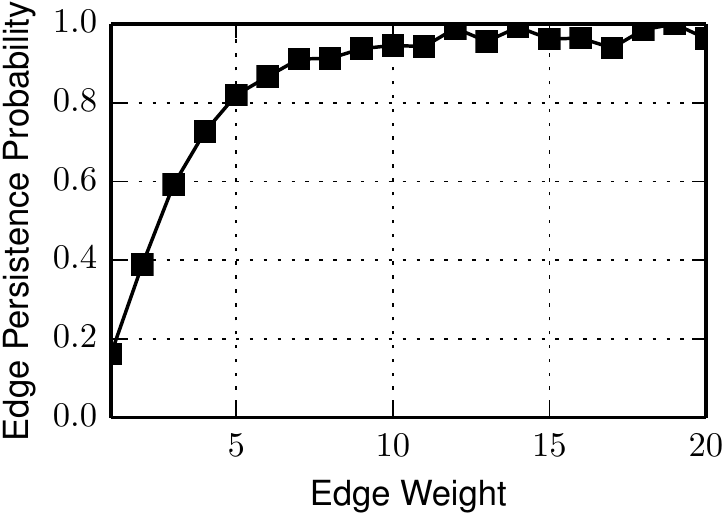}
        \caption{Probability that an edge will persist in the next network snapshot as a function of the edge's weight for New York.}
        \label{fig:weightpersists}
\vspace{-1.5mm}
\end{figure}
\begin{figure}[t]
        \centering
        \includegraphics[width=0.78\columnwidth]{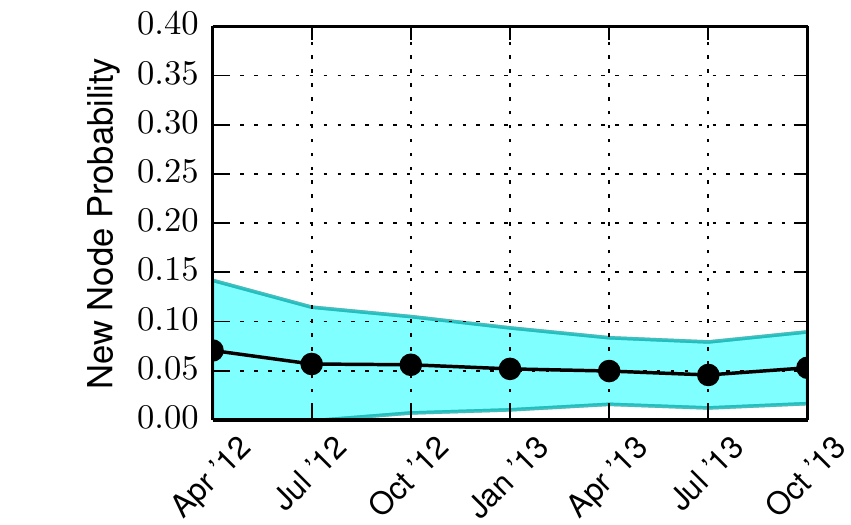}
        \caption{Probability of observing a new node in the next network snapshot.}
        \label{fig:newnodeprob}
\vspace{-1.5mm}
\end{figure}
\begin{figure}[t]
        \centering
        \includegraphics[width=0.78\columnwidth]{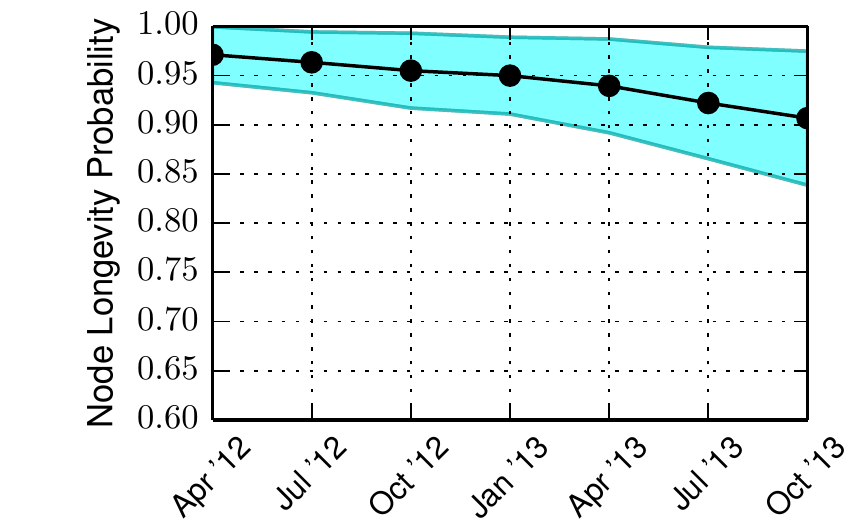}
        \caption{Node longevity across temporal network snapshots.}
        \label{fig:nodelongevity}
\vspace{-1.5mm}
\end{figure}
We observe that this probability is just above $30\%$ with values dropping towards $20\%$ over subsequent months. The small portion of stable links that propagate through time, roughly $20\%$, corresponds to high-weight edges representative of regular traffic patterns like commuting.  We will delve deeper into the functional role of places in the network in Section~\ref{sec:analysis}, where we observe that most high-weight edges are linking to travel hubs (airports, train stations etc.).
Figure~\ref{fig:weightpersists}, demonstrates that an edge's longevity directly relates to its weight. 
We report the probability that an edge will persist in the following snapshot as a function of the edge's weight defined as:
\begin{equation}
P_{e}(w) = \frac{|\{e \in E^{t} \cap E^{t+1}: weight(e) \geq w \}|}{|\{e \in E^{t} : weight(e) > w\}|}
\end{equation}
where $weight(e)$ is the function that returns the weight of an edge $e$. An edge with weight equal to $2$ has a probability very close to $0.4$ of being observed in the next time window, with the same probability doubling to a value very close to $0.8$
when the weight is equal to $5$. It should be noted that there is a possibility, albeit small, that high-weight edges may not re-appear in the future. Large events or transport disruptions could alter the flow of movement of urban populations~\cite{georgiev2014}, e.g. when a large crowd is moving to the park for a music festival. This can lead to the opportunistic 
appearance of high-weight edges that may not persist over long periods of time.
\vspace{-3.5mm}
\paragraph*{\textbf{Node persistence over time}}
We now investigate whether places persist over longer periods of time. As we did in 
the case of network links, we formulate the \textit{new node} and \textit{node longevity} probabilities
and observe how these evolve across several temporal snapshots.
Similarly to the new edge probability definition, the new node probability, defined as 
\begin{equation}
P_{u} = \frac{|V^{t+1} - V^{t}|}{|V^{t+1}|}
\end{equation}
for the set of nodes in $V^{t}$ during period $t$, is approximately $5\%$, as shown in Figure~\ref{fig:newnodeprob}. These results suggest that connectivity in cities emerges over a very stable base of venues with the constant generation of new links formed due to differences in users' mobility patterns. 
We also consider the persistence probability of nodes formalized as:
\begin{equation}
P_{v,n} = \frac{|V^{t} \cap V^{t+1} \dots \cap V^{t+n}|}{|V^{t}|}
\end{equation}
We observe that places seen in a snapshot, not only persist to the following one as implied by Figure~\ref{fig:newnodeprob}, but as shown
in Figure~\ref{fig:nodelongevity} they are active in the network with a probability greater than $95\%$ over many subsequent months with the value remaining above $90\%$ on average even after two years.

Taken together, these results paint a picture of a highly volatile network in terms of the edge generation process. Unlike online social networks, such as Facebook, where the friendships being formed among users tend to persist across time, in place networks the majority of links are constantly evolving and often fleeting, existing only for a short time period.  This observation motivates the construction of link prediction models that can track their evolution over time.  An improved understanding of the mechanisms behind network evolution is crucial in order to provide more intelligent location-based services that can adapt to the ever-changing patterns of cities and offer more tailored recommendations and advertisements based on a user's location and expected mobility patterns.
Prior to devising predictive models for link prediction in place networks, we investigate their topological properties in the following section.

\section{Properties of Place Networks}
\label{sec:analysis}
\begin{table*}
\centering
\begin{tabular}{|c||c|c|c|c|c|c|c|c|c|c|}
\hline
City & $|V|$ & $|E|$ & $C$ & $C_{r}$ & $D$ & $D_{r}$ & $d$ & $d_{r}$ & $\langle k \rangle$ & $r$ \\
\hline
Saint Petersburg & $9292$ & $278099$ & $0.20$ &  $0.08$  & $5.83$  & $5.91$ & $3.30$ &  $3.25$  &  $42.57$  &  $-0.05$  \\
\hline
Moscow & $8962$ & $168945$ & $0.19$ &  $0.07$  & $6.25$  & $6.00$ & $3.21$ &  $3.37$  &  $30.96$  &  $-0.05$   \\
\hline
Sao Paulo & $8643$ & $66110$ & $0.17$ &  $0.04$  & $6.83$  & $6.25$ & $3.67$ &  $3.68$  &  $18.01$  &  $-0.05$  \\
\hline
New York & $8156$ & $145671$ & $0.18$ & $0.07$ & $5.91$  & $5.25$ & $3.12$ & $3.14$ &  $40.99$  &  $-0.07$  \\
\hline
Kuala Lumpur & $7656$ & $56035$ & $0.19$ &  $0.05$  & $6.41$  & $6.00$ & $3.45$ &  $3.43$  &  $23.93$  &  $-0.06$  \\
\hline
Istanbul & $7389$ & $60790$ & $0.14$  & $0.02$ & $ 10.00$ & $ 7.50 $ &$4.66$ & $4.20$  &  $10.50$  &  $+0.05$ \\
\hline
Tokyo & $7327$ & $36627$ & $0.23$ &  $0.07$  & $7.58$  & $7.16$ & $3.79$ &  $3.79$  &  $10.40$  &  $-0.09$  \\
\hline
Bangkok & $6986$ & $33827$ & $0.15$ &  $0.04$  & $7.41$  & $6.58$ & $3.88$ &  $3.74$  &  $15.58$  &  $-0.00$  \\
\hline
Singapore & $6825$ & $30384$ & $0.14$ &  $0.02$  & $8.08$  & $7.08$ & $4.02$ &  $3.89$  &  $14.95$  &  $-0.00$  \\
\hline
Jakarta & $4645$ & $10776$ & $0.08$ &  $0.00$  & $10.83$  & $9.25$ & $5.45$ &  $5.07$  &  $5.84$  &  $+0.05$ \\
\hline
\end{tabular}
\caption{Statistics and network properties for a set of $10$ cities in the first three months of 2013. For each network we report the number of nodes $|V|$, number of edges $|E|$, mean clustering coefficient $C$, median network diameter $D$, mean shortest path $d$, average degree $\langle k \rangle$ for the undirected network versions and assortativity $r$. We denote the statistics of the corresponding null models with the subscript
$r$ where appropriate.}
\label{tab:foursquare_properties}
\vspace{-2.5mm}
\end{table*}
In Section~\ref{sec:evolution}, we have shown how place networks are similar to other networks in terms of densification patterns. We now ask whether this similarity manifests also with regard to the more elaborate properties of 
networks. We aim to answer the following question:
\textit{Do place networks in cities have similar structural properties to other social and technological networks that have been empirically investigated in the past or are they fundamentally different?} 

\begin{figure}
    \centering
            \includegraphics[width=0.8\columnwidth]{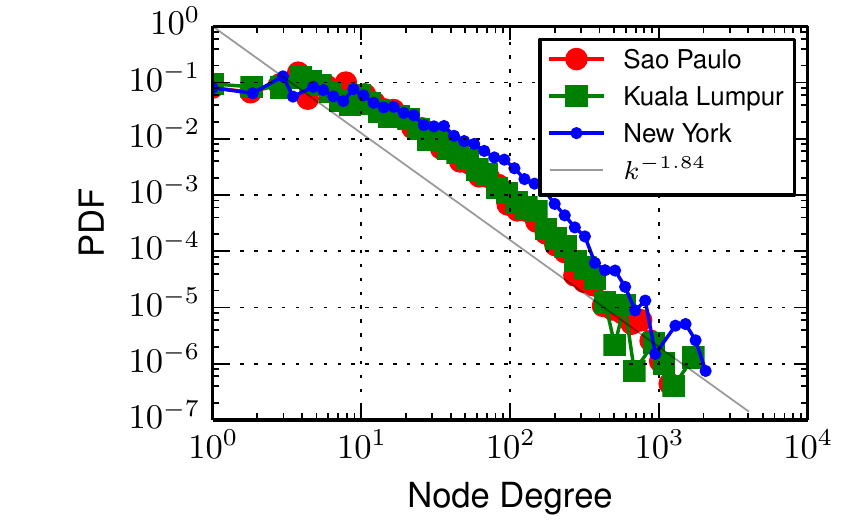}
            \caption{Probability density of the degree distribution in a temporal network snapshot for three cities and a linear fit to the plotted data on a log-log scale.}
            \label{fig:degrees}
\vspace{-1.5mm}
\end{figure}

\begin{figure}
    \centering
            \includegraphics[width=0.8\columnwidth]{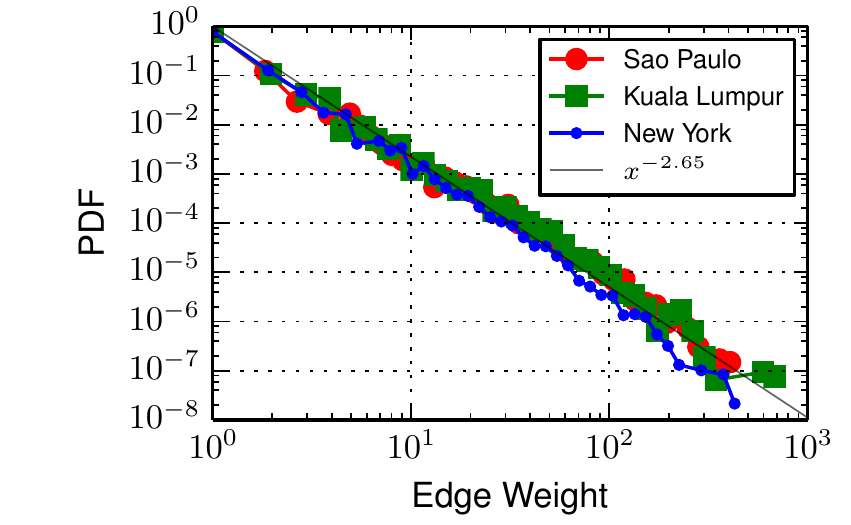}
            \caption{The edge weight distribution is more broad compared to the distribution of
            degrees shown in Figure~\ref{fig:degrees}.}
            \label{fig:weights}
\vspace{-1.5mm}
\end{figure}

\begin{figure}
    \centering
    \includegraphics[width=0.46\textwidth]{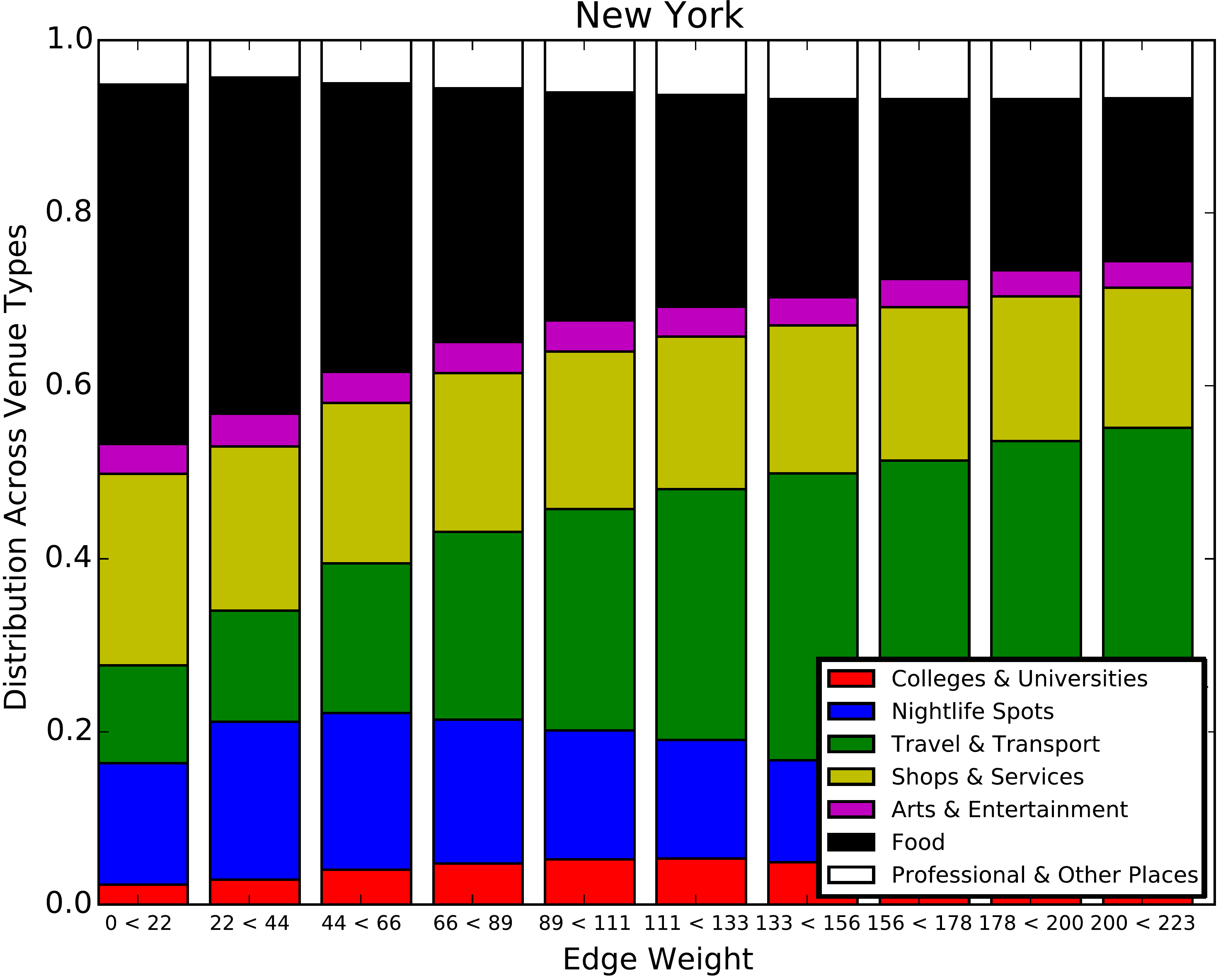} 
    \caption{Edge weight distribution by place category in New York with Travel \&  Transport in green
    color and Food in black.
    }
    \label{fig:typeweights}
\vspace{-1.5mm}
\end{figure}

Topological network properties such as the degree distribution, community structure or small-world behavior are known to have important implications for the functionality of real-world networks, including their robustness and information spreading.
Moreover, the existence of regularities in the network topology, associated with mechanisms driving link formation, is a central ingredient in algorithms for link recommendation, as seen in online social networks~\cite{adamic2003friends,liben2007link}. 
In Table~\ref{tab:foursquare_properties}, we summarize the statistics of the place networks built from the check-ins observed in the first three months of $2013$ for $10$ cities in our dataset.
\vspace{-2.5mm}
\paragraph*{\textbf{Network heterogeneity and the functional role of places in the urban domain}}
Many real world networks, including the world wide web and social networks, are 
known to exhibit a heavy-tailed degree distribution, often approximated by a power-law of the form
$P(k) \propto k^{-\beta}$. The latter is associated with a strong heterogeneity in the connectivity of the system, as a vast majority of nodes are poorly connected, while a few nodes play the role of hubs and inter-connect a large number of neighbors. 
As we show in Figure~\ref{fig:degrees}, place networks also exhibit a heavy-tailed degree distribution.  Using 
the maximum likelihood parameter estimation method~\cite{clauset2009power}, we fit the distribution, and find a mean power-law exponent $\beta$ equal to $1.84$ with a standard deviation of $0.09$ across $100$ cities. A similar relationship holds for the edge weight distribution in the network, shown in Figure~\ref{fig:weights}. The exponent $2.65\pm0.05$ indicates the existence of strong links, associated with a large flow of users, even on the order of several hundreds, between certain places. 
In order to interpret the strong heterogeneities present in the system, we exploit semantic information about Foursquare venue types. 
Figure~\ref{fig:typeweights}, illustrates the distribution of edge weights stratified by the underlying place categories (e.g. 'food', 'nightlife', etc.) for New York City. Food places (e.g restaurants, sandwich places, coffee shops etc.) dominate the sets of low-weight nodes, but are progressively replaced by transportation hubs as weight increases. This observation is in agreement with the fact that food places are the largest set in terms of number of places, i.e. that correspond to a mean fraction $0.4 \pm 0.05$ of the venues observed in a city, whereas travel places are dominant in terms of their proportion of check-ins, $0.13 \pm 0.07$, despite their fairly small number of venues (mean $0.04 \pm 0.02$). Both types of venues match their traditionally perceived role as functional units in the urban setting: that is, a diverse pool of food establishments operating alongside the presence of a few primary transportation hubs that handle most of the citizen and commuter mobility. While other place types co-exist 
in this environment, the dominance of food and travel spots, in terms of number of nodes 
and high weight links respectively, is remarkable. This potentially reflects two fundamental requirements 
of modern cities as part of the ever-intensifying urbanization process in modern development. First, the need for efficient urban mobility facilitated by transportation hubs, and second, the necessity for the presence of myriad food spots, spread around the city, in order to support human populations with 
vital resources. 

Note that, as the distribution of edge weights is well fitted by a power law, there is a vast number of edges that only occur once or twice. By definition, the latter do not survive long and tend to appear as new events in the network evolution,  as we discussed in the previous section (see Figure~\ref{fig:newedgeprob}). 
The fact that the distribution of node degrees is less peaked at small values 
(the probability is approximately uniform for degrees smaller than $10$), implies that places are more persistent than edges, as we observed  in Figure~\ref{fig:newnodeprob}.
\vspace{-2.5mm}
\paragraph*{\textbf{Triadic closure and small-world property}}
A list of the main network statistics for a sample of cities in the dataset is given in Table~\ref{tab:foursquare_properties}. 
We first note that the vast majority (almost $99\%$) of the nodes belong to 
the network's giant connected component $N_{GC}$.
For each city, we also present a randomized network (null model) where links are randomly rewired, preserving the number of nodes, edges and the original degree distribution of the places in the network. This way we are able to assess
the significance of the network measurements as compared to the corresponding random network.

Triadic closure is a central mechanism of social network evolution, typically measured with the average clustering coefficient $C$
\begin{equation}
C = \frac{1}{|V|}\sum_{u \in V} c_u
\end{equation}
where $c_u$ is the fraction of closed triangles between nodes connected to node $u$, also known as the local clustering coefficient. 
$C$ has been calculated for the undirected version of the place network, leading to a mean clustering coefficient value $\bar{C} = 0.20$ with a standard deviation $0.06$ across the full set of cities. The corresponding average value for the null models was significantly lower, $\bar{C_{r}} = 0.07 \pm 0.03$. 
The exact mechanisms leading to a high density of triangles in place networks may be different from those of social networks, yet it is notable that this property consistently holds  in the former class of networks. 
Let us also note that place networks are embedded in space, and that spatial embedding is a plausible mechanism leading to triadic closure ~\cite{scellato2011socio}.
The strong connections between social network topology and human mobility patterns~\cite{scellato2011exploiting, cho2011friendship, wang2011human, liben2005geographic} are also expected to make social networks and place networks share similar patterns.

Next, we focus on the distance between nodes, in terms of the number of hops between them
in the place network.
We report a mean shortest path $\bar{d} = 3.35 \pm 0.52$ across cities and a mean diameter $\bar{D}=6.35 \pm 1.46$, when the values of the randomized versions are $\bar{d_r} = 3.33 \pm 0.46$ and $\bar{D_r} = 5.93 \pm 1.19$ respectively. 
The existence of short paths connecting places in the network, together with their high clustering coefficient, imply that place networks 
are \textit{small-world}.
The 
small-world property can have significant implications for multiple processes in urban systems, including the spread of epidemic disease or information propagation and rumor spreading in cities, as has been described
in the context of many other network systems~\cite{kleinberg}. Despite the relationship with online social networks identified so far, 
as we demonstrate next, place networks present non-social connectivity patterns too.

\paragraph*{\textbf{Analogies to the web graph}}
Because of their strong relationships to human mobility, place networks can be viewed
as \textit{spatial navigation systems}~\cite{lee2012exploring}. 
For this reason, and inspired by  recent works on virtual navigation of web users in online domains~\cite{west2012automatic,west2012human}, 
we now investigate whether place networks reproduce non-social topological properties, associated with information systems like the web graph. To do so, we focus on the notion of \textit{assortativity}~\cite{newman2002assortative}.
Social networks are known to present positive assortative mixing, that is a tendency for high-degree nodes to be connected with each other.  On the contrary, the world wide web 
presents a mixed assortativity trend
with hubs connecting to each other but also to low-degree nodes. In the case of place networks, we observe a very similar behavior with a mean value of $\bar{r}= -0.055 \pm 0.04$. This value is remarkably consistent  across the one hundred cities, and very close to that observed empirically in the world wide web graph ($r = -0.065 $) and in protein interaction networks ($r = -0.156$) ~\cite{newman2002assortative}.
The latter observation is reminiscent of recent models of cities as biological organisms~\cite{bettencourt2007growth}. 

The assortativity scores in place networks can be explained by their polycentric, hierarchical structure (\cite{roth2011structure}), where two dominant connectivity patterns emerge: transport hubs (high degree nodes) connect to each other through the transportation system to facilitate commuter transit and, at the same time, hubs interact with a plethora of low degree nodes, associated with services, such as food and nightlife places, as users move to nearby places for refreshments and entertainment. This process is also reminiscent of \textit{authorities} and \textit{hubs} in the web, that is pages with authoritative content and others that connect to many of them.


\section{Link Prediction}
\label{sec:models}
The empirical analysis of place networks in Section~\ref{sec:evolution} has revealed that
a large number of new links is being generated steadily over time. 
Despite the large turnover of the links present in the system, the networks exhibit stable topological patterns, as described in Section~\ref{sec:analysis}.  
Our objective now is to exploit network structure together with mobility information about the temporal dynamics of places, in order to predict where edges will appear in a future time period. 
We manifest this goal by proposing a new gravity model in Paragraph~\ref{sec:gravity}. We also formalize
a set of network and mobility models (Paragraph~\ref{sec:features}) and present a host of supervised learning algorithms (Paragraph~\ref{sec:supervised}) that we use as a testbed to assess the model's efficacy.
\vspace{-2.5mm}
\paragraph*{\textbf{Problem formulation}}
Given a graph representing the place network of a city $G^{t} = (V, E^{t}) $ during a period $t$, the goal is to predict the edges $E^{t+1}$ that appear in the network during the next temporal snapshot $t+1$. The problem essentially consists of ranking pairs of nodes $(i,j)$ according to a numeric score $r_{ij}$ estimating the likelihood of an edge appearing
between a source node $i$ and a destination node $j$, with $i,j \in V$. 
Link prediction has been a popular problem  
in recent years and numerous models have been proposed to determine $r_{ij}$: from unsupervised predictors primarily exploiting node topology in the network~\cite{liben2007link}, to supervised learning algorithms that integrate multiple signals synchronously~\cite{lichtenwalter2010new}. 
\paragraph*{\textbf{Challenges}}
\begin{figure}
    \centering
    \includegraphics[width=0.75\columnwidth]{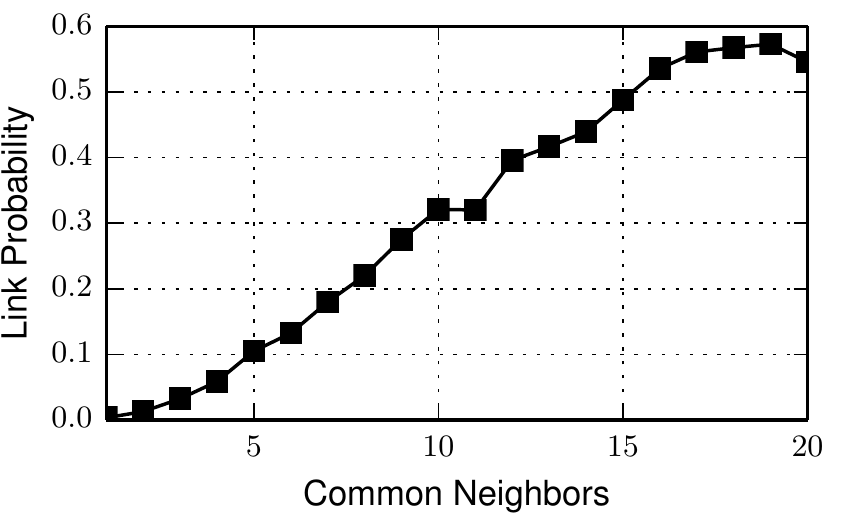}
    \caption{The probability of link formation in the next temporal network snapshot 
    as a function of the common neighbors shared by two nodes in the network, shown 
    for Chicago.}
    \label{fig:edgeprobacommoneighs}
\vspace{-1.5mm}
\end{figure}
The introduction of the link prediction problem in the domain of urban place networks brings new challenges
that the candidate prediction models need to address. 
First, the realization of the \textit{short term temporal dynamics} in user mobility due to the 
existence of periodicities and diurnal patterns in human activity during a week.  
For example, a corporate office is more likely to connect to a coffee shop on a Wednesday morning as opposed 
to a Saturday night. 
Second, the ability to cope with the \textit{long term temporal dynamics} characterizing the system, 
since there is significant novelty in terms of edges in the network over time; activity in the system drift
as users alter their mobility patterns during urban exploration or due to seasonal changes.
Third, \textit{geography} is known to play an important role in connectivity in urban environments~\cite{noulas2012tale}.
As a consequence, link prediction models deployed in cities should integrate information on the distance between places effectively. 
\subsection{Network formation and human mobility }
We view an urban place network as an entity shaped by the interplay
of two primary forces: its current \textit{network form}, as seen in a given window of observation, and human movement, which is the generative force of connectivity in the system. Our underlying assumption is that this feedback loop between network and mobility forces is key to the network's evolution. In the following, we define a number of models, associated with different link formation mechanisms, which we exploit in the link prediction problem. 
\vspace{-3.5mm}
\paragraph*{\textbf{Network predictors}}
Let us first introduce \textit{network} models for the link prediction task.  
We are guided by the common properties between place networks and social networks, discussed in Section~\ref{sec:analysis}.
Being also inspired by popular algorithms for link prediction in online social networks, we introduce a set of predictors which are based on two key factors in link generation: triadic closure and node centrality.
\begin{itemize}
\vspace{-1.5mm}
\item{\textbf{Triadic closure mechanics.}} A standard predictor to determine whether two nodes in a network will connect is the number of common neighbors they share. We thus define the \textbf{CommonNeighbors} feature for two venues $i$ and $j$ as $|\Gamma_i  \cap \Gamma_j |$.
In the above formulation, the greater in number the common neighbors of two venues, the more likely they are to interact in the future through a direct user transition. We verify this assumption in Figure~\ref{fig:edgeprobacommoneighs} where we see a significant increase in the probability of inter-place connectivity as the number of common neighbors for a pair of places grows.
Jaccard's similarity coefficient has been proposed~\cite{liben2007link} as an improvement to this indicator by taking into account, additionally, the size of the neighborhoods of the two places. We refer to this model as \textbf{NeighborOverlap} and define it as $\frac{| \Gamma_i  \cap \Gamma_j |}{| \Gamma_i \cup \Gamma_j |}.$

Furthermore, we define the \textbf{AdamicAdar} indicator that depends on the number of common neighbors of two places, but also penalizes those common neighbors that have high degrees:
\begin{equation}
\sum_{ z \in \Gamma_i \cap \Gamma_j} \frac{1}{\log(|\Gamma_{z}|)}
\end{equation}
This measure is known as a very efficient predictor in online social networks~\cite{adamic2003friends}. In Section~\ref{sec:evaluation}, we will show that this is also the case for place networks. For this reason, we also integrate it into the new gravity model we present in Paragraph~\ref{sec:gravity}.
\item{\textbf{Node centrality metrics.}}
In addition to the previous predictors based on triangles, we also focus on predictors purely based on the importance of the nodes. A basic centrality measure is degree centrality, and it is incorporated into  the feature \textbf{DegreeProduct} as $|\Gamma_i|.|\Gamma_j|$, which can be generalized by taking into account the direction of the links, \textbf{InOutDegreeProduct} as $|\Gamma_i^{+}|.|\Gamma_j^{-}$|.
Finally, we also use a non-local measure of centrality, PageRank~\cite{page1999pagerank}, denoting the score of place $i$ as $rw(i)$, and define the \textbf{PlaceRank} indicator $rw(i).rw(j)$.
\end{itemize}
\paragraph*{\textbf{Mobility predictors}}
\label{sec:features}
We now introduce a set of spatial and mobility information signals related to places. 
\begin{itemize}
\item{\textbf{Static mobility metrics.}}
Geographic distance is well-known to have an impact on human mobility ~\cite{brockmann2006scaling, gonzalez2008understanding}, and its effect has been evaluated in the case of location-based social networks~\cite{cho2011friendship, noulas2012mining}. For this reason, we use the \textbf{GeoDistance} predictor that ranks two candidate venues $i$ and $j$ according to their geographic distance $d(i,j)$ measured in kilometers. 
The closer two venues are, the higher their position will be in the prediction list, with the implicit assumption being that nearby places are more likely to form a link at a future time.
Next, we define a \textbf{Popularity} feature that ranks venue pairs according the product of their popularity $c_ic_j$, where $c_i$ is simply the sum of check-ins at place $i$. 
The strong effect of \textit{geographic distance} and \textit{popularity} on the probability of a link existing for a pair 
of places is shown in Figure~\ref{fig:distancepopedge}. 
\item{\textbf{Mobility dynamics.}}
\begin{figure}
    \centering
    \includegraphics[width=0.7\columnwidth]{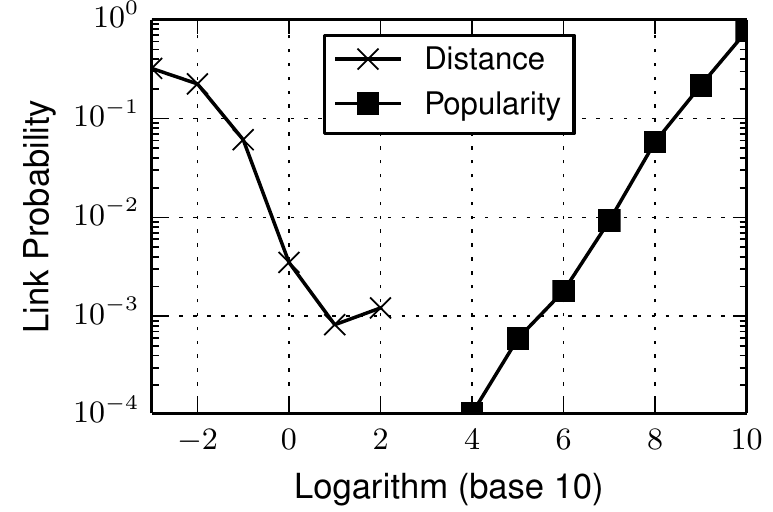}
    \caption{The probability of link formation in the next temporal network snapshot 
    as a function of the geographic distance between two places and the product
    of their popularities in New York.}
    \label{fig:distancepopedge}
\vspace{-1.5mm}
\end{figure}
\begin{figure*}
\centering
\begin{subfigure}[b]{0.9\columnwidth}
\includegraphics[width=0.9\columnwidth]{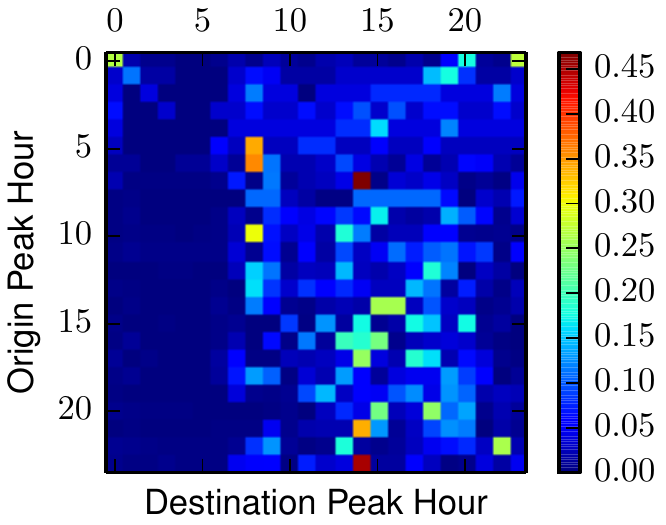}
\caption{Weekdays}
\label{fig:wdays}
\end{subfigure}
\centering
\begin{subfigure}[b]{0.9\columnwidth}
\includegraphics[width=0.9\columnwidth]{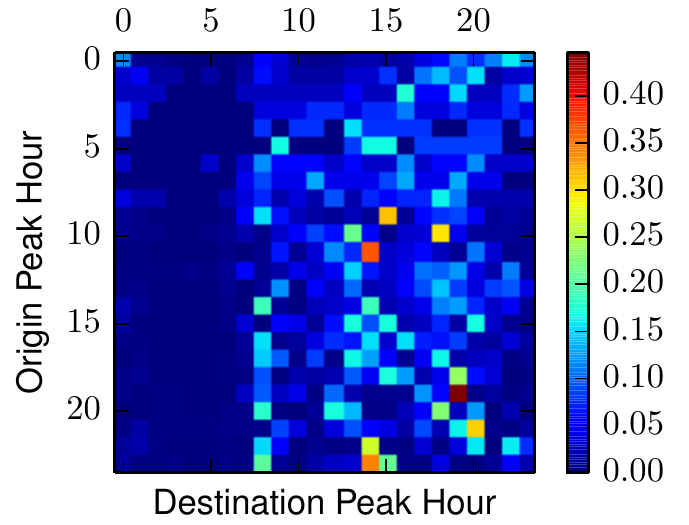}
\caption{Weekends}
\label{fig:wends}
\end{subfigure}
\caption{Diurnal venue interaction patterns aggregating over weekdays (top) and weekends (bottom) for the city of London. On the y-axis (rows) we note the peak hour of the origin venue, and on the x-axis (columns) that of a destination. Each point refers to the density of transitions between venues in a given hour slot.}
\label{fig:timeedges}
\end{figure*}
The popularity of Foursquare venues does not remain static over time however. It is constantly changing with strong diurnal and weekly patterns due to corresponding variations in urban human movement and activities. Since we are able to capture the precise time of each check-in we can explicitly model these fluctuations. We define the temporal similarity between two places $i$ and $j$ as the cosine similarity between their $T$-dimensional check-in frequency vectors equal to:
\begin{equation}
 cos(\vec{c_i}^\tau, \vec{c_j}^\tau)~:~\tau \in \{1,T\}
\end{equation}
$T$ can be set equal to $24$ or $168$ to capture daily and weekly cycles of activity respectively. We denote the diurnal and weekly similarity models as \textbf{DiurnalSim} and \textbf{WeeklySim} respectively.
Both metrics are based on the assumption that two venues are more likely to connect when they are visited by mobile users during similar hours. 
\end{itemize}
The importance of place synchronization is shown in Figure~\ref{fig:timeedges}, where we visualize the relationship between venue interaction frequencies and venue peak hours (peak hour is simply the hour when the number of check-ins of a venue maximizes). Plotting on the y-axis the peak hour of an origin venue we can see the probability of forming an edge with a destination venue that peaks at a certain hour of the day. 
Venues that are active in the morning are relatively more likely to connect to venues that peak in the morning too,
with this relationship holding considering different temporal intervals through day and night.
In the following paragraph, we fuse information about the mobility dynamics seen here with the network features presented above, to
introduce a new gravity model that realizes the characteristics of modern mobility datasets.
\subsection{Gravity Models}
\label{sec:gravity}
The use of gravity models to reproduce human migration patterns roots back to the seminal work of Ravenstein~\cite{ravenstein1885laws}. Different variants have been developed, initially in transportation research and urban planning in the 50s and 60s~\cite{carrothers1956historical,niedercorn1969economic} and more recently when studying large-scale mobility patterns measured in census or cellular data~\cite{brockmann2006scaling, gonzalez2008understanding}. 
Inspired by Newton's law of gravity, gravity models assume that the flux between two regions $i$
and $j$ takes the form $P_{ij} \propto k_i k_j f(d_{ij})$ where $f(d_{ij})$ is a deterrence function decreasing with the geographic distance $d_{ij}$ between $i$ and $j$. $k_i$ is a measure of the region's \textit{mass}, its attractiveness. Taking a standard form for the deterrence function $f(d) \sim d^{-\beta}$, we propose in this section two versions of gravity models: a classic formulation, as defined in the literature,
and a new gravity model that incorporates the importance of network structure and temporal synchronization between regions.
Note that in the present domain places are used as proxy for regions.
\begin{itemize}
\item{\textbf{A static formulation of gravity.}}
The notion of mass, in gravity models, depends on the nature of the system under scrutiny.
In the case of fluxes of mobility between cities, for instance, it is common to use their population as a proxy.
In our classic version of the \textit{gravity model}, we therefore use the total popularity of place as its mass, and define a  \textbf{Gravity} index as:
\begin{equation}
\frac{\displaystyle c_ic_j}{\displaystyle d(i,j)^{\beta}}
\label{equation:gravity}
\end{equation}
This model is essentially a combination of the \textbf{Popularity} and \textbf{GeoDistance} features, defined in the previous paragraph and whose effect on place connectivity has been illustrated in Figure~\ref{fig:distancepopedge}. 
\item{\textbf{A dynamic gravity model.}}
Modern mobile datasets have the advantage, as compared to old-fashioned data concerning mobility, e.g. based on censuses, to include a fine temporal resolution. Our analysis of place networks suggests that the temporal patterns of node activity play an important role in link formation. This observation motivates the incorporation of temporal information into the gravity model in order to account for  the inherently dynamic aspects of human mobility. Similarly, our results concerning triadic closure also suggest the incorporation of network information in the modeling approach. 

For this reason, we propose a new gravity model, called \textbf{DynamicGravity}, that brings together
three fundamental aspects for link prediction: network structure, mobility dynamics and geography. 
Formally, the model is defined as:
\begin{equation}
\frac{\displaystyle a_{ij} \sum_{\tau=1}^{T} c_i(\tau)^{+} c_j(\tau)^{-}}{\displaystyle d(i,j)^{\beta}}
\end{equation}
with $T=168$ and $c_i(\tau)^{+}$ noting the out-strength of place $i$, that is the sum of weights of all out-going edges, during hour $\tau$. 
Equivalently, $c_i(\tau)^{-}$ accounts for the in-strength at a given time instance. $a_{ij}$ is the \textbf{AdamicAdar} score between two places defined above. 
The model
aims at capturing in a principled way the most important information signals in the data.

First, the importance of a place in the network in terms of  its \textit{strength}. Second the \textit{directionality of the edges} generated by a place given that places act as sources or sinks in the network. Third, the \textit{temporal dynamics and periodicities} of these forces as they fluctuate in time. Finally the potential connections between nodes driven by triadic closure.
While these components favor connectivity between important nodes in the network (hubs), geographic distance acts in this context as a balancing factor and biases connectivity towards nearby 
places. This allows for the hubs to connect towards the plethora of low-degree nodes that are present all over the city and the mixed assortativity trends seen in Section~\ref{sec:analysis}
are also realized.
\end{itemize}

At this point, note that there is a possibility for two places not to share any common neighbors in the network. These places are still able to connect in the future. 
To accommodate this potential, we extend the original formulation of the Adamic-Adar~\cite{adamic2003friends} model to the form $\sum_{ z \in \Gamma_i \cap \Gamma_j} \frac{1}{\log(|\Gamma_{z}|)} + 1.0$. 
The unit value is added to represent the connectivity of all nodes in the network towards an imaginary node $i_{m}$ and thus model the possibility of all nodes in the system to interact at the network level. 
Finally, note that when no information about time is available ($T=1$), there are no common neighbors between two nodes $(a_{ij}=1$) and there is no information about edge directionality, the equation of the dynamic gravity model
falls back to the setup of the original, static gravity model formalized in Equation~\ref{equation:gravity}.

\subsection{Supervised learning methods}
\label{sec:supervised}
Supervised learning methods have been recently hypothesized to provide effective solutions to the link prediction problem~\cite{lichtenwalter2010new}, compared to unsupervised predictors such as those presented in the previous paragraphs. Supervised learning algorithms are aware of 
the class imbalance (ratio of positive and negative instances) in the system, and unlike their unsupervised counterparts, are able to operate across multiple differentiating class boundaries and capture the inter-dependency of many variables. Even if a single variable is considered (e.g. distance)
and multiple decision boundaries exist, a supervised learning method can learn that more complex relationship.
Finally, as more data flows in the system through time and more training labels become available, supervised learning models are only expected to improve in terms of accuracy.
\\
\textbf{Training methodology:} We build a training dataset on the previous network snapshot, using positive and negative labels for connected pairs and disconnected pairs of nodes
respectively. We have experimented both with a balanced and an imbalanced training dataset case where the class distribution is preserved, choosing the latter as it yielded better experimental results in all cases. For each pair of places $i$ and $j$, we construct input feature vectors $\mathbf{x_{ij}}$ formed by all network and mobility features introduced in Paragraph~\ref{sec:features}. Here, we employ three classifiers: \textit{logistic regression}~\cite{fan2008liblinear} (applied with l$2$ regularization) and the ensemble learning methods \textit{random forests}~\cite{breiman2001random} (optimized here with $50$ trees of maximum depth $50$) and \textit{gradient boosting}~\cite{friedman2001greedy} (with $100$ tree estimators of maximum depth $10$).

\section{Evaluation}
\label{sec:evaluation}
\paragraph*{\textbf{Experimental Setup}}
We assess the ranking performance of all models in the link prediction task by means of Area Under the Curve (AUC) score. To extract the Receiver Operating Characteristic (ROC) curve we measure the True Positive Rate versus False Positive Rate ratio for varying decision thresholds on the ranked list of pairs of places in each city. Then the AUC score is calculated by measuring the area under the ROC curve and it provides an indication 
of the performance of an information signal in balancing the trade-off between precision and recall. A predictor that ranks place pairs randomly would yield 
a ROC curve matching the diagonal line, $y=x$, and hence an AUC score $0.5$.
We make use of a more informed baseline, named \textbf{EdgeWeight}, that simply ranks pairs of nodes
according to the their edge weight in the previous temporal snapshot (if an edge does not exist the score is $0$). We test the performance of all ranking strategies in a realistic temporal cross-validation setting by training on the first three months of 2013 and testing on the following temporal snapshot in 
the same year. We summarize the performance of all models in Table~\ref{table:auc} showing their average score in terms of AUC
and standard deviations across the one hundred cities in the dataset. 
\vspace{-1mm}
\\
\paragraph*{\textbf{Results}}
Focusing on the network models, we observe that the 
model which exploits \textit{edge directionality information} of the nodes, \textbf{InOutDegreeProduct},
performs best with an AUC score $0.875$ improving predictability over the \textbf{DegreeProduct} that ignores directionality (AUC=$0.862$)
and scoring equally with the PageRank adaptation, \textbf{PlaceRank}. The former demonstrates that there are places in the network that tend to behave as sources or sinks by generating or absorbing mobility flows of users.
The models that do the next best in terms of AUC performance are those based on common neighbors between places; \{textbf{CommonNeighbors} scores AUC=$0.824$ and \textbf{AdamicAdar} attains a very similar score, AUC=$0.822$.

We next evaluate the class of models that are built on geographic or mobility information about places. While geographic distance explains to some extent the connectivity between places and performs better than the naive predictor \textbf{EdgeWeight}, it is clearly being outperformed by the \textbf{Popularity} predictor
that captures the significance of places in the network in terms of how frequently they are visited by Foursquare users. Ranking performance improves further
when the temporal visitation patterns of users at places is taken into account as the \textbf{WeeklySim} model suggests. As implied by the significant improvement (AUC=$0.851$ versus AUC $0.774$) over \textbf{DiurnalSim} which employs information about diurnal mobility patterns only,
\textit{the temporal synchronization between places across weekdays and weekends matters}. 
\begin{table}[h!]
\centering
\begin{tabular}{|cccc|}
\hline
\textbf{Model} & \textbf{$AUC$} &  \textbf{$\pm$} &  \\
\hline
\hline
\centering
Network & & & \\
\hline
\texttt{\textbf{DegreeProduct}} & 0.862 & 0.020&\\
\texttt{\textbf{InOutDegreeProduct}} & 0.875 & 0.021&\\
\texttt{\textbf{NeighborOverlap}} & 0.803 & 0.038&\\
\texttt{\textbf{CommonNeighbors}} & 0.824 & 0.048&\\
\texttt{\textbf{AdamicAdar}} & 0.822 & 0.050&\\
\texttt{\textbf{EdgeWeight}} & 0.626 & 0.028&\\
\texttt{\textbf{PlaceRank}} & 0.875 & 0.022&\\
\hline
Geo-Mobility & & &\\
\hline
\texttt{\textbf{GeoDistance}} & 0.702 & 0.047&\\
\texttt{\textbf{Popularity}} & 0.768 & 0.039&\\
\texttt{\textbf{DiurnalSim}} & 0.774 & 0.040&\\
\texttt{\textbf{WeeklySim}} & 0.851 & 0.025&\\
\hline
Multi-variate & & &\\
\hline
\texttt{\textbf{Gravity}} & 0.811 & 0.041&\\
\texttt{\textbf{DynamicGravity}} & 0.905 & 0.019&\\
\texttt{\textbf{RandomForests}} & 0.864 & 0.071&\\
\texttt{\textbf{GradientBoost}} & 0.687 & 0.104&\\
\texttt{\textbf{LogisticReg}} & 0.792 & 0.047&\\
\hline
\end{tabular}
\vspace{0.2in}
\caption{Mean \emph{AUC} and standard deviation scores across cities for all features and models.}
\label{table:auc}
\end{table}
Looking across all prediction methods, we note that the dynamic gravity model, \textbf{DynamicGravity}, outperforms by a clear margin all models in the list. Notably, it raises the performance of the classic gravity formulation by almost ten points (AUC=0.905
vs 0.811). This shows how \textit{the temporal dynamics of user mobility at places combined with information about the connectivity patterns and structure of the place network, formed by the trajectories of mobile users, can significantly improve predictability over mobility models that utilize solely static information}. 
The dynamic version of the gravity model we present here essentially fuses the best information
signals from the mobility and network classes above (\textbf{InOutDegreeProduct, AdamicAdar} and \textbf{WeeklySim}) as well as integrating geographic distance as a factor. It effectively captures the fact that places not only behave as sources or sinks in the network, but also that the way they adhere to these roles in the system changes dynamically over time. For example, one would expect 
a school, or a corporate office, to be a sink node in the morning, that becomes a source
when it terminates operations later in the day. At the same time a transportation hub in the area may
follow an inverse pattern, being a source in the morning and a sink in the evening, in accordance with 
local commuter mobility trends. 
\vspace{-1mm}

Finally, the model outperforms not only the static gravity version and all network and mobility models, but also supervised learning algorithms that have previously been shown to excel in link prediction in other domains~\cite{lichtenwalter2010new}. 
Besides, the latter require special care in terms of training and parameter optimization. 
This can be costly in many realistic deployment scenarios that demand crisp and accurate responses, 
a frequent case in mobile application settings.
The parameter $\beta$ in the case of the \textbf{DynamicGravity} model has been set equal to $1$ for all cities. Its simple formulation 
has allowed it to generalize well without overfitting in a very volatile network environment, where link formation can be influenced 
by seasonal drifts, changes in individual user mobility patterns or even large social events that can alter the mobility flows of user populations in the urban domain~\cite{georgiev2014}.


\section{Discussion and Related Work}
\label{sec:related}
In this work, we have investigated the properties of urban place networks formed by the check-in patterns of millions of Foursquare users across a large set of $100$ cities around the globe.
We have shown that these networks exhibit many of the well-known properties of other social technological networks, and that the growth patterns of these networks are characterized by a dynamic edge generation process over a relatively stable set of nodes.  

Understanding the way venues interact in the urban domain by means of user mobility patterns can support existing applications and pave the way for new ones. The science of placing a new retail facility or venue, for instance, in an already established urban network of places would benefit from precise information about network connectivity in the local area and its underlying dynamics~\cite{jensen2006network,karamshuk2013geo}. Seeing cities as dynamically evolving networks is in line with recent works that propose the exploitation of network-based techniques to understand modern urban systems~\cite{bettencourt2013origins, bettencourt2007growth, batty2012smart} 

Using network analysis techniques to understand how cities grow and evolve 
may not be a novelty on its own, yet
most research in this domain has focused on street network analysis~\cite{crucitti2006centrality,porta2006network} or the analysis of transportation networks~\cite{roth2011structure, Louf}. 
The novelty of our approach is based on positioning real world places at the spotlight of network analysis
using as input Foursquare's venue database.
While the network of places has been a fundamental element in related works in location-based services, either for the detection of neighborhoods as in the case of the Livehoods project~\cite{cranshaw2012livehoods} or to perform venue recommendations~\cite{noulas2012mining}, its properties have not been studied empirically in the past. In that respect the analysis we conduct can benefit future works where the place network becomes important.

Our results have implications for the development of modern mobile applications too. 
The image of a highly volatile network in terms of the edge generation process in particular motivates the construction of link prediction models that can track their evolution over time. Unlike in online social networks, such as Facebook, where the friendships being formed among users tend to persist across time, in place networks the majority of links are constantly evolving and often fleeting, existing only for a short time period. A better understanding of the mechanisms behind network evolution is crucial in order to provide more intelligent location-based services that can adapt to the ever-changing patterns of cities and offer more tailored recommendations and advertisements based on the user's location and expected mobility patterns.


\section*{Acknowledgments}
Anastasios Noulas acknowledges the support of EPSRC through Grant GALE (EP/K019392) and the Foursquare family for the great times in New York.

\small
\bibliographystyle{plain}
\bibliography{biblio}
\end{document}